\documentclass[twocolumn,preprintnumbers,amsmath,amssymb,superscriptaddress,10pt]{revtex4}
\usepackage{amssymb}
\usepackage{graphicx,color}
\usepackage{bm}
\usepackage{ulem}

\begin{document}

\title{Dynamical scheme for hadronization with first-order phase transition}

\author{Bohao Feng}
\affiliation{Department of Physics, Tsinghua University and Collaborative Innovation Center of Quantum Matter, Beijing 100084, China}

\author{Zhe Xu \footnote{xuzhe@mail.tsinghua.edu.cn}}
\affiliation{Department of Physics, Tsinghua University and Collaborative Innovation Center of Quantum Matter, Beijing 100084, China}

\author{Carsten Greiner}
\affiliation{Institut f$\ddot{u}$r Theoretische Physik, Johann Wolfgang Goethe-Universit$\ddot{a}$t Frankfurt, Max-von-Laue-Strasse 1, 60438 Frankfurt am Main, Germany}

\begin{abstract}
We present a dynamical scheme for hadronization with first-order confinement phase transition.
The thermodynamical conditions of phase equilibrium, the fluid velocity profile, and the
dissipative effect determine the macroscopic changes of the parton volume and the corresponding 
hadron volume during the phase transition. The macroscopic volume changes are the basis 
for building up a dynamical scheme by considering microscopic transition processes from partons
to hadrons and backwards. The established scheme is proved by comparing the numerical results
with the analytical solutions in the case of a one-dimensional expansion of a dissipative fluid
with Bjorken boost invariance. The comparisons show almost perfect agreements, which demonstrate
the applicability of the introduced scheme.
\end{abstract}

\maketitle

\section{Introduction}
\label{sec1:intro}
The relativistic heavy-ion collisions provide an opportunity in the laboratory to investigate
QCD matter under extreme conditions of high temperature, high density, and strong
electromagnetic field. Data taken in experiments at the Relativistic Heavy Ion Collider
(RHIC) \cite{Arsene:2004fa,Back:2004je,Adams:2005dq,Adcox:2004mh} and at the Large Hadron
Collider (LHC) \cite{Aamodt:2010pa,Aad:2010bu,Chatrchyan:2011sx} indicate the transient
existence of a quark-gluon plasma (QGP), which then undergoes phase transitions and
gradually merges into a large number of hadrons. We are interested in the dynamical process
of the phase transition, which is essentially needed, in order to have a complete physical
picture of relativistic heavy-ion collisions and to understand phenomena found at RHIC and
LHC. In particular, the dynamical description of the phase transition could determine the
contribution of gluons in the buildup of the collective flow of hadrons, which has not been
intensively studied so far. In  quark coalescence models \cite{Lin:2002rw}, which have been
employed to explain the quark number scaling behavior in the hadronic elliptic flows found
at RHIC \cite{Abelev:2008ae}, gluons are not explicitly considered. 

Another motivation concerns the viscous effect during the phase transition. In viscous
hydrodynamical calculations
\cite{Luzum:2008cw,Dusling:2009df,Song:2010aq,Schenke:2010rr,Niemi:2011ix}
the shear viscosity to the entropy density ratio ($\eta/s$) of the parton-hadron mixture
during the phase transition is set to be constant. However, this treatment is only an 
assumption, since there is no evidence for the equal $\eta/s$ of partons and hadrons at
the phase transition. The dynamical description of the phase transition would determine
the real viscous corrections to the thermal distribution functions of each hadron species
\cite{Molnar:2014fva} and would examine the applicability of the Cooper-Frye prescription \cite{Cooper:1974mv} used in viscous hydrodynamical calculations.

Since the dynamical description for the phase transition from first principle is at
present an unsolved problem, we have to content ourselves with modeling, which allows
exploring related phenomena in an articulated way. In this article we will introduce
a dynamical scheme for the confinement phase transition of first order.

The purpose of this article is conceptional. We consider, for simplicity, the
transition from gluons to pions. The gluons that we concern are soft particles, which
build up the bulk of the medium. We do not discuss the hadronization of gluon jets.
Also, we do not discuss the transition from gluons
to glueballs \cite{Stoecker:2015zea}, which is a sharp first-order phase transition
\cite{Svetitsky:1982gs}. We {\it assume} that the hadronization from gluons to pions
is a first-order phase transition. Although this contradicts the fact that the QCD
transition at zero baryon chemical potential is a crossover
\cite{Aoki:2005vt,Bazavov:2009zn}, the condition of phase equilibrium that keeps the
temperature and chemical potential of gluonic and pionic phase equal and constant
during the first-order phase transition will prove numerical implementations.
Simulating the crossover phase transition needs the correct implementation of the 
equation of state (EoS) from the lattice QCD calculations and is a future project. 
The present work can be seen as an attempt to describe a first-order phase transition
between two phases with different degrees of freedom. It is the first step towards
a full scheme describing the first-order phase transition from quarks and gluons to
mesons and baryons at a finite baryonic chemical potential. Adding quarks and more
hadron species into the scheme is more complicated, but in line with the present
implementation, and will be shown in a forthcoming paper.

The numerical implementation of hadronization that we introduce is a further extension
of the existing parton cascade BAMPS (Boltzmann Approach of Multi Parton Scatterings)
\cite{Xu:2004mz}, which describes the pre-equilibrium stage, the thermalization, and 
the hydrodynamical evolution of quarks and gluons produced in ultrarelativistic
heavy-ion collisions. The dynamical hadronization scheme will serve as an interface
between BAMPS and hadronic transport model, which will be developed next.
BAMPS is a numerical solver of the kinetic Boltzmann equations for on-shell quarks
and gluons by using test particles
to represent phase space distribution functions of quarks and gluons. Interactions
of quarks and gluons are simulated by the stochastic interpretation of the transition
rates of scattering processes. The numerical implementation of transitions from gluons
to pions, which will be presented in this article, has the same means as used in BAMPS
for interactions of quarks and gluons. We will show that the effective probabilities
of the microscopic processes for transitions from gluons to pions are entirely
determined by the thermodynamical feature of the phase transition, the viscosity of
the QCD matter, and the velocity profile of the hydrodynamical expansion. Our numerical
implementation is different from the hadronization procedures used in transport models
such as AMPT (A multiphase transport model) \cite{Lin:2004en}, PHSD 
(Parton-Hadron-String Dynamics) \cite{Cassing:2008sv}, etc.

The article is organized as follows. In Sec. \ref{sec2:theory} we derive the volume
change of gluons and pions during the phase transition, based on the conditions of
phase equilibrium at the first-order phase transition and hydrodynamical equations.
With this we establish a dynamical scheme transferring gluon matter to pion matter
during the first-order phase transition in Sec. \ref{sec3:model}.
In Sec. \ref{sec4:bjorken} the analytical formulas of the gluonic volume fraction,
number, energy, and entropy density are derived in the case of a one-dimensional
expansion with Bjorken boost invariance, in order to prove the numerical
implementations by comparing the analytical solutions with numerical results shown
in Sec. \ref{sec6:results}. Before doing the comparisons, we present details of
numerical implementations and setups in Sec. \ref{sec5:setup}. Finally we summarize
and give an outlook in Sec. \ref{sec7:summary}.

\section{The EoS and the first-order phase transition in
a gluon-pion mixture}
\label{sec2:theory}
For the EoS of gluons we employ the standard MIT bag model \cite{Chodos:1974je}.
The pressure and energy density are
\begin{eqnarray}
\label{pg}
&& P_g=\frac{1}{3} (e_g - 4B)=n_g T_g -B \,, \\
\label{eg}
&& e_g=3n_g T_g +B \,,
\end{eqnarray}
where $n_g$ denotes the gluon number density and $T_g$ is the temperature.
For the bag constant we use $B^{1/4}=0.23 \mbox{ GeV}$.
The pion system is considered as an ideal gas. We neglect pion's rest mass
for simplicity. The pressure and energy density of massless pions are then
\begin{eqnarray}
\label{ppi}
&& P_\pi=\frac{1}{3} e_\pi =n_\pi T_\pi \,, \\
&& e_\pi=3n_\pi T_\pi \,,
\end{eqnarray}
where $n_\pi$ denotes the pion number density and $T_\pi$ the temperature.
Here we have ignored the quantum Bose enhancement \cite{Xu:2014ega} 
of gluons and pions and regarded them as Boltzmann particles.

For the first-order phase transition, both EoS of gluons and pions are matched
to each other via the Gibbs condition
\cite{Rischke:1995mt,Sollfrank:1996hd,Kolb:2000sd},
\begin{equation}
\label{gibbs}
P_g=P_\pi \equiv P_c \,, \ \ T_g=T_\pi \equiv T_c \,, \ \ 
\mu_g=\mu_\pi \equiv \mu_c \,.
\end{equation}
$\mu_g$ and $\mu_\pi$ are the chemical potential of gluons and pions, respectively,
which are defined by
\begin{equation}
\label{chemicalpotential}
e^{\frac{\mu_i}{T_i}}= \frac{n_i}{n_i^{eq}}\,,
\end{equation}
where $i$ stands for $g$ or $\pi$. $n_i^{eq}$ is the particle number density
in thermal equilibrium,
\begin{equation}
n_i^{eq}=\frac{d_i}{\pi^2} T_i^3 \,,
\end{equation} 
where $d_g=16$ and $d_\pi=3$ are the degeneracy factor of gluons and pions,
respectively. 

Now we consider the confinement phase transition in an expanding QCD matter.
Suppose $V$ is the volume of an expanding element in its local rest frame at proper
time $\tau$. During the phase transition the volume of pions is increasing, while
the volume of gluons is decreasing. We denote $V_g$ and $V_\pi$ as the volume of
gluons and pions. The fraction of the gluon phase to the mixture is then
$f_g=V_g/V=V_g/(V_g+V_\pi)$. The total particle number and energy density are
\begin{eqnarray}
\label{nm}
&&n_m = n_g^c f_g + n_\pi^c (1-f_g) \,, \\
\label{em}
&&e_m = e_g^c f_g + e_\pi^c (1-f_g) \,,
\end{eqnarray}
where $n_g^c$, $e_g^c$ ($n_\pi^c$, $e_\pi^c$) are the particle number and energy
density of gluons (pions) at the transition temperature $T_c$, respectively. From
the above equations for $n_m$ and $e_m$, and the EoS of gluons and pions it follows
\begin{equation}
\label{emp}
e_m+P_c=4n_m T_c \,. 
\end{equation}

In the following we derive the time dependence of $f_g$ in a local region under
the Gibbs condition (\ref{gibbs}). In our dynamical scheme gluons hadronize smoothly
into pions. We do not consider spinodal instabilities \cite{Chomaz:2003dz}, which
lead to fluctuations in the baryon density \cite{Steinheimer:2012gc,Li:2016uvu},
for instance. It would be possible to introduce spinodal instabilities when adding
quarks and baryons in our scheme and incorporating the mean field into Vlasov
term of the Boltzmann equation.

During a time step $d\tau$ the considered volume element is expanded to $V+dV$. 
The volume of gluons is decreased to $V_g+dV_g$, while the volume of pions is
increased to $V_\pi+dV_\pi$. $dV_g$ is negative. Thus, $dV_\pi=dV-dV_g$ is larger
than $-dV_g$. The volume changes indicate that $-n_g^c dV_g$ gluons are confined
into $n_\pi^c dV_\pi$ pions and an energy of a amount of $-e_g^c dV_g$ has to be
redistributed to the pionic and gluonic phase in order to maintain the Gibbs
condition Eq. (\ref{gibbs}).

For a hydrodynamic system, its energy density changes according to the
hydrodynamical equation \cite{Muronga:2003ta},
\begin{equation}
\label{vhydro}
De=-(e+P)\nabla_\mu U^\mu +\pi^{\mu\nu} \nabla_{<\mu} U_{\nu>}\,,
\end{equation}
where $U^\mu$ is the fluid four-velocity and $\pi^{\mu \nu}$ is the shear tensor.
Symbols in the above equation are defined as follows:
\begin{eqnarray}
&&D=U^\mu \partial_\mu \,, \\
&&\nabla^\mu=\Delta^{\mu\nu} \partial_\nu \,, \\
&&\Delta^{\mu\nu}=g^{\mu\nu}-U^\mu U^\nu \,,\\
&&A^{<\mu\nu>}=\left [ \frac{1}{2} \left ( \Delta^\mu_\sigma \Delta^\nu_\tau
+\Delta^\nu_\sigma \Delta^\mu_\tau \right ) -\frac{1}{3} \Delta^{\mu\nu} \Delta_{\sigma\tau}
\right ] A^{\sigma\tau} \,.
\end{eqnarray}
In Eq. (\ref{vhydro}) the heat transfer is neglected and the bulk pressure is zero,
since here we consider systems of massless particles. The right hand side of
Eq. (\ref{vhydro}) can be written as $-(e+P_{eff}) \nabla_\mu U^\mu$ by introducing
an effective pressure $P_{eff}=P + \tilde \pi$, where
\begin{equation}
\label{shearpressure}
\tilde \pi=- \frac{\pi^{\mu\nu} \nabla_{<\mu} U_{\nu>}}{\nabla_\mu U^\mu} \,.
\end{equation}
For a pure one-component system, the kinetic energy in the rest frame of an expanding
volume element decreases by $dE=-P_{eff} dV$ due to the work done by the effective
pressure. Thus, the temperature decreases too.

In order to hold the Gibbs condition (\ref{gibbs}) during the phase transition,
there must be energy influxes into the gluonic and pionic phase, which compensate
the energy loss of $dE_g=-(P_c+\tilde \pi_g) f_g dV$ and
$dE_\pi=-(P_c+ \tilde \pi_\pi) (1-f_g) dV$ in the gluonic and pionic phase, respectively.
All these energies should come from the transition energy $-e_g^c dV_g$.
After subtracting $dE_g$ and $dE_\pi$ from $-e_g^c dV_g$, the remaining energy is the
energy of newly produced pions and must be equal to $e_\pi^c dV_\pi$, in order to keep
the temperature of pions as $T_c$. The energy balance reads
\begin{equation}
\label{energy}
-e_g^c dV_g -( P_c+\tilde \pi_g) f_gdV -(P_c+ \tilde \pi_\pi) (1-f_g)dV = e_\pi^c dV_\pi \,.
\end{equation}
Inserting the EoS of gluons, Eqs. (\ref{pg}) and (\ref{eg}), into the left-hand side
of the energy balance (\ref{energy}) gives
\begin{eqnarray}
\label{energy1}
&&-(3n_g^c T_c +B)dV_g -( n_g^c T_c -B +\tilde \pi_g) f_gdV 
\nonumber \\
&&-(P_c+ \tilde \pi_\pi) (1-f_g)dV \nonumber \\ 
&=&-3n_g^c T_c dV_g +B(-dV_g +f_g dV) -( n_g^c T_c +\tilde \pi_g) f_gdV 
\nonumber \\
&&-(P_c+ \tilde \pi_\pi) (1-f_g)dV \,.
\end{eqnarray}
The second term on the right-hand side of the above equation, which is proportional
to the bag constant, is the latent heat, $dE_{lat}$, provided by the bag pressure
in volume $V$ during time $d\tau$. Then the terms in the energy balance Eq. (\ref{energy})
are rearranged to
\begin{eqnarray}
\label{energy2}
-3n_g^c T_c dV_g +dE_{lat}&=&e_\pi^c dV_\pi + ( n_g^c T_c +\tilde \pi_g) f_gdV \nonumber \\
&& +(P_c+ \tilde \pi_\pi) (1-f_g)dV \,.
\end{eqnarray}
We see that the kinetic energy of hadronizing gluons together with the absorbed
latent heat cover the energy of produced pions with $T_c$ and the loss of kinetic
energies of gluons and pions due to the work done by the effective pressure.

Putting $dV_\pi=dV-dV_g$ in the energy balance Eq. (\ref{energy}) we obtain
\begin{equation}
\label{dvg}
dV_g=-\frac{e_\pi^c+P_c+\tilde \pi_m}{e_g^c-e_\pi^c}dV
\end{equation}
with $\tilde \pi_m=\tilde \pi_g f_g+ \tilde\pi_\pi(1-f_g)$.
$dV$ can be determined according to the identity
\begin{equation}
\label{hydroexp}
\frac{1}{V} \frac{dV}{d\tau}=\nabla_\mu U^\mu \,.
\end{equation}
From the definition of $f_g$ and Eqs. (\ref{dvg}) and (\ref{hydroexp}), we have
\begin{eqnarray}
\label{dfg}
\frac{df_g}{d\tau}&=&\frac{1}{V} \frac{dV_g}{d\tau}
-f_g\frac{1}{V}\frac{dV}{d\tau} \nonumber \\
&=&\left [-\frac{e_\pi^c+P_c+ \tilde \pi_m}{e_g^c-e_\pi^c}
-f_g \right ] \nabla_\mu U^\mu \,,
\end{eqnarray}
which can be solved to obtain the time dependence of $f_g$. Once we know
$U^\mu$ and $\pi^{\mu\nu}$ from transport or hydrodynamic calculations, we can
determine $dV_g$ and $f_g$. In addition, the latent heat can be expressed as
\begin{equation}
\label{condx}
dE_{lat}= B(-dV_g +f_g dV)= \frac{e_m +P_c +\tilde \pi_m}{e_g^c-e_\pi^c} B dV \,.
\end{equation}

We notice that Eq. (\ref{dvg}) can be derived in a pure mathematical way.
For that we first differentiate the energy density in Eq. (\ref{em}) with respect
to $\tau$ and equate this with the hydrodynamical equation (\ref{vhydro}) to get
$df_g/d\tau$. We then use the first identity of Eq. (\ref{dfg}) to obtain $dV_g$,
which is found to be identical to Eq. (\ref{dvg}). This consistence confirms the
correct dynamical picture of the first-order phase transition near equilibrium.

Equation (\ref{dvg}) is indeed an important result, which shows quantitatively
how the transition between gluons and pions proceeds and is a basic equation for
establishing a microscopic transport scheme for the first-order phase transition.
Although Eq. (\ref{dvg}) has been derived for a transition from gluons to pions in
an expanding volume element, it is also valid for a transition from pions to gluons
in a contracting volume element, where $dV$ and $\nabla_\mu U^\mu$ are negative. 
In this case the volume element gains energy from the surrounding medium.
The energy balance in Eq. (\ref{energy2}) can be reinterpreted that the sum of
the energy from the transition $-e_\pi^c dV_\pi$ and that from the surrounding medium
$-(n^c_g T_c+\tilde \pi_g)f_g dV$ and $-(P_c+\tilde \pi_\pi)(1-f_g) dV$ is equal to
the sum of the kinetic energy of newly produced gluons  $3n_g^c T_c dV_g$ and
the released latent heat $-dE_{lat}$.

From Eq. (\ref{dvg}) we see the viscous effect on the phase transition. For a perfect
fluid, where $\tilde \pi_m=0$, $|dV_g/dV|$ is a constant, whereas for a viscous fluid
$|dV_g/dV|$ is time dependent and is smaller (larger) than that for $\tilde \pi_m=0$
in a transition from gluons to pions (from pions to gluons), since $\tilde \pi_m$ is
negative (positive) in an expanding (a contracting) system [see Eq. (\ref{shearpressure})].
The different behavior of $|dV_g/dV|$ in transitions from gluons to pions and backwards
is due to the fact that the process of the phase transition with non-zero viscosity
is irreversible.

Moreover, in an expanding system $e_\pi^c+P_c+\tilde \pi_m$ could be negative for
large $|\tilde \pi_m|$, so that $dV_g$ would become positive, which cannot describe
the phase transition from gluons to pions, where $dV_g$ should be negative. This indicates
that for large dissipation the first-order phase transition cannot occur. Quantitative
statements about an upper limit of the shear viscosity will be made elsewhere.
We mention that it seems that there is no such upper limit of the shear viscosity
for the phase transition from pions to gluons in a contracting system, since $\tilde \pi_m$
is positive.

Finally, a nonzero shear viscosity will increase the total entropy during the
phase transition. This important feature will be realized in the to be introduced 
dynamical scheme of hadronization. Before we proceed, the entropy density
is given by
\begin{equation}
\label{entropydensity}
s_i=\frac{e_i+P_i-\mu_i n_i}{T_i}=\left ( 4- \frac{\mu_i}{T_i}
\right ) n_i \,,
\end{equation}
where $i$ stands for $g$ or $\pi$. During the phase transition the total entropy
density is
\begin{equation}
\label{sm}
s_m = s_g^c f_g + s_\pi^c (1-f_g) =\left (4-\frac{\mu_c}{T_c} \right ) n_m \,,
\end{equation}
where $s_g^c$ and $s_\pi^c$ are the entropy density of gluons and pions at the
transition temperature $T_c$.

\section{The dynamical scheme for hadronization}
\label{sec3:model}
In the rest of the article we consider only the phase transition from gluons
to pions in expanding systems.

Using Eq. (\ref{dvg}) we find that the difference between the number of gained
pions and the number of lost gluons in the volume element $V$ during $d\tau$ is
\begin{equation}
\label{diff}
n_\pi^c dV_\pi - (-n_g^c dV_g)=-\frac{\tilde \pi_m}{4T_c} dV \,,
\end{equation}
which is non-negative, since $\tilde \pi_m \le 0$ from Eq. (\ref{shearpressure}).
This indicates that for an ideal fluid the number of gained pions is the same as
that of lost gluons, while for a viscous fluid the number of gained pions is
larger than that of lost gluons, which increases the total entropy. Therefore,
we in principle need number-changing processes, such as two gluons go to three
pions, $g+g \to \pi +\pi +\pi$, to implement hadronization in a viscous fluid.
We will see later that a part of the latent heat provides an additional energy
to the three pions, so that the temperature and chemical potential of pions
keep constant.

For the phase transition from gluons to pions we consider the following processes:
$g+g \to \pi+\pi+\pi$, $g+g\to \pi+\pi$, and back reactions $\pi+\pi+\pi\to g +g$
and $\pi+\pi \to g+g$. Here we hide the charge of pions, which could be noted
explicitly as $g+g \leftrightarrow \pi^+ + \pi^- +\pi^0$, 
$g+g \leftrightarrow \pi^0 + \pi^0 +\pi^0$, $g+g \leftrightarrow \pi^+ +\pi^-$,
and $g+g \leftrightarrow \pi^0 +\pi^0$. The probabilities of the occurrence of
these processes could be tuned to obtain the same yield of all kind of pions. 
We have to note that the consideration of these microscopic processes is not
from the first principle but is necessary to realize the macroscopic volume change
according to Eq. (\ref{dvg}) and to maintain the Gibbs condition (\ref{gibbs}). 
Therefore, in principle one could consider other processes. The processes
we have considered are the simplest one can think of.

When gluons hadronize into pions in the process $g+g \to \pi+\pi+\pi$ and 
$g+g\to \pi+\pi$, besides the total kinetic energy of gluons, an amount of energy
from the bag pressure (latent heat), will be involved in the total energy of pions.
Therefore, the average energy of each produced pion is larger than that of the
lost gluons, which is $3T_c$.
We denote the ratio of the total energy of the final pions over the total kinetic
energy of the initial gluons by $x$, which is larger than $1$.
We will show later that the determination of the ratio $x$ [see Eq. (\ref{px})] 
corresponds to the latent heat [see Eq. (\ref{condx})]. It is obvious that the total
kinetic energy is not conserved in the processes $g+g \to \pi+\pi+\pi$ and
$g+g\to \pi+\pi$. Since transitions with momentum and kinetic energy conservation
have been numerically implemented in a standard routine, we {\it amplify}
the momentum (also the kinetic energy) of each gluon by $x$ before performing the
transitions to pions by using the standard routine. One can easily prove that the
factor $x$ is Lorentz invariant. 

Since the latent heat has been involved in $g+g \to \pi+\pi+\pi$ and
$g+g\to \pi+\pi$ according to Eq. (\ref{condx}), in back reactions
$\pi+\pi+\pi \to g+g$ and $\pi+\pi \to g+g$ the total momentum as well as the
total kinetic energy are conserved. We allow only those back reactions to occur,
if pions are newly produced from $g+g \to \pi+\pi+\pi$ and $g+g\to \pi+\pi$.
Thus, on average, each gluon coming from back reactions has a larger energy than
$3T_c$. This mimics the energy transfer from the pionic phase to the gluonic phase, 
in order to compensate for the energy loss of gluons due to the hydrodynamical expansion.
We involve back reactions in processes $g+g \to \pi+\pi+\pi \to g^* + g^*$,
$g+g \to \pi+\pi+\pi \to g^* + g^*+\pi$, and $g+g \to \pi+\pi \to g^* + g^*$,
where three pions or two pions are regarded as intermediate states and $g^*$
denotes outgoing gluons with a higher averaged energy than that of initial gluons.
Numerically we implement $g+g \to g^* + g^*$ and  $g+g \to g^* + g^*+\pi$ directly
and do not specify intermediate states explicitly.

Now we derive the probability that a process $g+g \to \pi+\pi+\pi$,
$g+g \to \pi+\pi$, $g+g \to g^* + g^*+\pi$, or $g+g \to g^* + g^*$ occurs,
denoted by $P_{23}$, $P_{22}$, $P_{23b}$, and $P_{22b}$, respectively.
For simplicity, these probabilities are assumed to be independent on the momenta
of particles involved in the processes. Therefore, the number of lost gluons and
gained pions in volume $V$ during $d\tau$ relate to the probabilities $P_{23}$,
$P_{22}$, and $P_{23b}$ as follows:
\begin{eqnarray}
\label{nglost}
&&\frac{1}{2} N_g (N_g-1) \left (2 P_{23}+ 2 P_{22} \right )=-n_g^c dV_g\,, \\
\label{npgain}
&&\frac{1}{2} N_g (N_g-1) \left (3 P_{23}+ 2 P_{22} + P_{23b} \right )=n_\pi^c dV_\pi \,,
\end{eqnarray}
where $N_g=n_g^c f_g V$ is the gluon number in volume $V$. Suppose the number
of $g^*$ from the back reactions is $dN_{g^*}$; then we have
\begin{equation}
\label{nggain}
\frac{1}{2} N_g (N_g-1) \left (2 P_{23b}+ 2 P_{22b} \right )=dN_{g^*} \,.
\end{equation}

The total kinetic energy of initial gluons in each transition process is
$6T_c$ ($3T_c$ for each) on average. As introduced before, we enhance
the kinetic energy of initial gluons by a $x$ factor, in order to include
the latent heat. The total energy involved in each transition process is then
$6T_c x$ on average, while the latent heat per process is $6T_c(x-1)$.
The total involved latent heat in volume $V$ during $d\tau$ relates to the sum
of the probabilities of all the transition processes as well as the factor $x$,
\begin{eqnarray}
\label{elat}
&&\frac{1}{2} N_g (N_g-1) \left (P_{23}+ P_{22} +P_{23b} + P_{22b} \right ) 6T_c(x-1) \nonumber \\
&=&dE_{lat}=\frac{e_m +P_c +\tilde \pi_m}{e_g^c-e_\pi^c} B dV \,.
\end{eqnarray}
The second identity is due to Eq. (\ref{condx}). 

In the process $g+g \to \pi+\pi$ the energy of each pion is $3T_c x$ on average,
while it is $2T_c x$ in the processes $g+g\to \pi+\pi+\pi$ and $g+g\to g^*+g^*+\pi$.
The average energy of each pion obtained from $g+g \to \pi+\pi$,
$g+g\to \pi+\pi+\pi$, and $g+g\to g^*+g^*+\pi$ should be larger than $3T_c$.
In other words, the total energy of these pions should be larger than 
$3T_c n_\pi^c dV_\pi=e_\pi^c dV_\pi$, because the energy excess over $e_\pi^c dV_\pi$
should cover the energy loss of all pions in volume $V$ due to the work done by the
effective pressure. This requirement leads to
\begin{eqnarray}
\label{epgain}
&&\frac{1}{2} N_g (N_g-1) \left (P_{23}+ P_{22}+ \frac{1}{3} P_{23b} \right )
6T_c x \nonumber \\
&=& e_\pi^c dV_\pi + (n_\pi^c T_c +\tilde \pi_\pi)(1-f_g) dV \,.
\end{eqnarray}

Analogously, the total energy of the gained gluons in the processes
$g+g\to g^*+g^*$ and $g+g\to g^*+g^*+\pi$ should be larger than $3T_c dN_{g^*}$,
because the excess should cover the energy loss of all gluons in volume $V$ due to
the work done by the kinetic pressure, which leads to
\begin{eqnarray}
\label{eggain}
&&\frac{1}{2} N_g (N_g-1) \left (P_{22b}+ \frac{2}{3} P_{23b} \right )
6T_c x \nonumber \\
&=& 3T_c dN_{g^*} + (n_g^c T_c +\tilde \pi_g)f_g dV \,.
\end{eqnarray}
We eliminate $dN_{g^*}$ by inserting Eq. (\ref{nggain}) and obtain
\begin{eqnarray}
\label{eggain1}
&&\frac{1}{2} N_g (N_g-1) \left [ P_{22b}(x-1) +  P_{23b} \left ( \frac{2}{3}x-1 \right )
\right ] 6T_c \nonumber \\
&=& (n_g^c T_c +\tilde \pi_g)f_g dV \,.
\end{eqnarray}

We notice that Eq. (\ref{eggain1}) plus Eq. (\ref{epgain}) minus $3T_c$ times
Eq. (\ref{nglost}) is equal to Eq. (\ref{elat}) by using the energy balance
Eq. (\ref{energy2}). This indicates that there are only four independent equations,
Eqs. (\ref{nglost}), (\ref{npgain}), (\ref{epgain}), and (\ref{eggain1}),
available for five unknowns, namely, four probabilities $P_{23}$, $P_{22}$, $P_{23b}$,
$P_{22b}$, and the factor $x$. One of five unknowns is a free parameter.
The determination of this free parameter should ensure that all the probabilities are
non-negative and $x$ is larger than $1$. We choose $P_{23}$ as the free parameter and
set it to be zero. With this choice all other transition probabilities are positive
and the factor $x$ is larger than $1$, as shown later in Fig. \ref{fig3-crossection}.
$P_{23}=0$ does not mean that there are no $g+g \to \pi+\pi+\pi$ processes,
but indicates that once such a process occurs, either three or two pions will go back
to two gluons, which are denoted by the processes $g+g \to \pi+\pi+\pi \to g^*+g^*$ or
$g+g\to \pi+\pi+\pi \to g^*+g^*+\pi$.

With $P_{23}=0$ we obtain $P_{22}$ directly from Eq. (\ref{nglost})
\begin{eqnarray}
\label{p22}
P_{22} &=& -\frac{n_g^c dV_g}{N_g(N_g-1)} \nonumber \\
&=& \frac{n_g^c(e_\pi^c +P_c + \tilde \pi_m)}{e_g^c-e_\pi^c}\frac{dV}{N_g(N_g-1)}
\nonumber \\ 
&=& \frac{n_g^c(e_\pi^c +P_c + \tilde \pi_m)}{e_g^c-e_\pi^c}
\frac{\nabla_\mu U^\mu V d\tau}{N_g(N_g-1)}
\end{eqnarray}
by using Eqs. (\ref{dvg}) and (\ref{hydroexp}).
Subtracting Eq. (\ref{nglost}) from Eq. (\ref{npgain}) gives
\begin{eqnarray}
\label{p23b}
P_{23b} &=&\frac{2}{N_g (N_g-1)} (n_\pi^c dV_\pi + n_g^c dV_g)
\nonumber \\
&=& -  \frac{\tilde \pi_m}{2T_c} \frac{\nabla_\mu U^\mu V d\tau}{N_g(N_g-1)} \,. 
\end{eqnarray}
To get the second identity we have used Eqs. (\ref{diff}) and
(\ref{hydroexp}).
Putting $P_{23b}$ and $P_{22}$ into Eq. (\ref{epgain}) we obtain
\begin{equation}
\label{px}
x=\frac{e_\pi^c dV_\pi + (n_\pi^c T_c +\tilde \pi_\pi)(1-f_g) dV}
{T_c(2n_\pi^c dV_\pi-n_g^c dV_g)} \,,
\end{equation}
where one may insert the ratios $dV_g/dV$ and $dV_\pi/dV=1-dV_g/dV$ from
Eq. (\ref{dvg}). Finally we get $P_{22b}$ from Eq. (\ref{eggain1}),
\begin{equation}
\label{p22b}
P_{22b}= \frac{(n_g^c T_c +\tilde \pi_g) f_g 
+(x-\frac{3}{2}) \tilde \pi_m}{3T_c (x-1)} \frac{\nabla_\mu U^\mu V d\tau}{N_g(N_g-1)} \,.
\end{equation}
With the derived probabilities $P_{22}$, $P_{23b}$, $P_{22b}$, and the factor $x$,
we can perform the corresponding transition processes stochastically in the same
manner introduced in BAMPS \cite{Xu:2004mz}.

For the phase transition from pions to gluons in contracting systems we can
analogously consider processes $\pi+\pi \to g+g$, $\pi+\pi \to \pi^*+\pi^*+g$,
and $\pi+\pi \to \pi^*+\pi^*$. The factor $x$ is in this case smaller than $1$,
because a latent heat will be released. The procedure of deriving the probabilities
and $x$ is same as that shown above.
 
Since $U^\mu$ and $\pi^{\mu\nu}$ can be extracted from the particle distributions
in transport calculations, our dynamical scheme for hadronization with the
first-order phase transition can in principle be applied for any systems.
In this article we will show a simulation in a particular case, where we consider
one-dimensional expansion with Bjorken boost invariance \cite{Bjorken:1982qr},
which is widely used to describe the space-time evolution of matter produced in
ultrarelativistic heavy-ion collisions. In this case the time evolution of the
phase transition can be calculated analytically, which we use to examine our
numerical implementations.

\section{The case of one-dimensional expansion with Bjorken boost invariance}
\label{sec4:bjorken}
In one-dimensional expansion with Bjorken boost invariance, the hydrodynamical
velocity is 
\begin{equation}
\label{flowvelocity}
U^\mu=\frac{1}{\tau}(t,0,0,z) \,.
\end{equation}
In the first-order theory of hydrodynamics, the shear tensor reads
\begin{equation}
\pi^{\mu\nu}=2\eta \nabla^{<\mu} U^{\nu>} \,,
\end{equation}
where $\eta$ is the shear viscosity. Then Eqs. (\ref{hydroexp}), (\ref{shearpressure}),
and (\ref{vhydro}) are reduced to
\begin{eqnarray}
\label{hydroexp2}
&& \frac{1}{V} \frac{dV}{d\tau} =\nabla_\mu U^\mu =\frac{1}{\tau} \,, \\
\label{shearpressure2}
&& \tilde \pi=-2\eta \frac{\nabla^{<\mu} U^{\nu>} \nabla_{<\mu} U_{\nu>}}{\nabla_\mu U^\mu}
=-\frac{4\eta}{3\tau} \,,\\
\label{detau}
&& \frac{de}{d\tau}=-\frac{e+P}{\tau}+\frac{4\eta}{3\tau^2} \,.
\end{eqnarray}
Using Eqs. (\ref{sm}), (\ref{emp}), and (\ref{detau}) we obtain the differential
equation for the time evolution of the entropy density during the phase transition
\begin{eqnarray}
\label{dsmdt}
\frac{ds_m}{d\tau}&=&\left ( 4- \frac{\mu_c}{T_c} \right )
\frac{dn_m}{d\tau} =\left ( 4- \frac{\mu_c}{T_c} \right )
\frac{1}{4T_c} \frac{de_m}{d\tau} \nonumber \\
&=&  -\frac{s_m}{\tau} + \left ( 1- \frac{\mu_c}{4T_c} \right )
\frac{4\eta_m}{3T_c \tau^2}
\end{eqnarray} 
with $\eta_m=\eta_g f_g+ \eta_\pi (1-f_g)$. $\eta_g$ ($\eta_\pi$) is the shear
viscosity of gluons (pions). Assuming that $\eta_m/s_m$ is a constant during
the phase transition, we solve Eq. (\ref{dsmdt}) and obtain
\begin{equation}
\label{smtau}
s_m(\tau)=s_g^c \frac{\tau_c}{\tau} e^{\frac{4a}{3T_c}
\left ( \frac{1}{\tau_c}- \frac{1}{\tau} \right )} \,,
\end{equation}
where $\tau_c$ is the time when the phase transition begins and
$a=(1-\mu_c/4T_c)\eta_m/s_m$. Thus, we get the gluonic fraction in the mixture
according to Eq. (\ref{sm}),
\begin{equation}
\label{fgtau_exp}
f_g(\tau)=\frac{s_m(\tau) - s_\pi^c}{s_g^c-s_\pi^c} \,.
\end{equation}
$f_g$ decreases from $1$ at $\tau_c$ to $0$ at $\tau_e$, which denotes the time
when the phase transition in the considered volume element is complete.
In addition, using Eqs. (\ref{sm}) and (\ref{emp}) we have
\begin{eqnarray}
\label{nmtau}
n_m(\tau)&=&n_g^c \frac{\tau_c}{\tau} e^{\frac{4a}{3T_c}
\left ( \frac{1}{\tau_c}- \frac{1}{\tau} \right )} \,, \\
\label{emtau}
e_m(\tau)&=&(e_g^c+P_c) \frac{\tau_c}{\tau} e^{\frac{4a}{3T_c}
\left ( \frac{1}{\tau_c}- \frac{1}{\tau} \right )}-P_c \,.
\end{eqnarray}

\section{Numerical implementations and setups}
\label{sec5:setup}
In this section we give details on numerical implementations and setups for
simulating the hadronization in a one-dimensional Bjorken expansion.
Since the main goal of this work is to present a dynamical scheme of hadronization
and to prove its applicability by comparing the numerical results with analytical
solutions, we consider only elastic scatterings among gluons or pions and assume
constant cross sections and the isotropic distribution of collision angles.
Under these assumptions we can easily tune the cross sections to have a constant
$\eta_m/s_m$ ratio. which is required to obtain analytical solutions; see 
Eqs. (\ref{smtau}) - (\ref{emtau}).

Elastic collisions among gluons or pions are simulated by employing the standard BAMPS
prescription. The collision probabilities \cite{Xu:2004mz} read
\begin{equation}
\label{probab}
P_i=v_{rel.} \frac{\sigma_i}{N_{test}} \frac{\Delta t}{f_i V_r} \,,
\end{equation}
where $i$ stands for either a process $g+g \to g+g$ or for $\pi+\pi\to \pi+\pi$,
and $\sigma_i$ is the respective cross section. $v_{rel.}$ denotes the relative
velocity of two incoming particles, and $V_r$ is the volume of a cell in the
computational frame. (Remember that $V$ is the cell volume in its local rest frame.)
$f_i$ is the gluon or pion fraction, which is $f_g$ or ($1-f_g)$. $\Delta t$ is the
time step in the computational frame, and $N_{test}$ is the number of test particles
per a real particle. 

For the isotropic distribution of collision angles the shear viscosity turns out
to be \cite{Huovinen:2008te,Wesp:2011yy,El:2012cr}
\begin{equation}
\label{eta}
\eta_i=\frac{6T_i}{5\sigma_i} \,.
\end{equation}
Then we can solve Eq. (\ref{detau}) and obtain the time evolution of the energy
density of gluons before the phase transition and that for pions after the phase
transition:
\begin{eqnarray}
\label{egtau}
e_g(\tau)&=&[e_g(\tau_0)-B] \left ( \frac{\tau_0}{\tau} \right )^{r_g} +B \,, \\
e_\pi(\tau)&=&e_\pi(\tau_e) \left ( \frac{\tau_e}{\tau} \right )^{r_\pi}  \,,
\end{eqnarray}
where $\tau_0$ is the initial time of the gluonic phase, and $r_g$ and $r_\pi$ are given by
\begin{eqnarray}
r_g&=&\frac{4}{3} - \frac{8}{15 n_g(\tau_0) \tau_0 \sigma_g} \,, \\
r_\pi&=&\frac{4}{3} - \frac{8}{15 n_\pi(\tau_e) \tau_e \sigma_\pi} \,.
\end{eqnarray}

For completeness we give the solutions of the time evolution of number density
and temperature, which is defined by the ratio of the kinetic energy density
over threefold of the number density,
\begin{eqnarray}
\label{ntau}
&& n_g(\tau)=n_g(\tau_0) \frac{\tau_0}{\tau} \,, \
n_\pi(\tau)=n_\pi(\tau_e) \frac{\tau_e}{\tau} \,, \\
\label{temp}
\label{ttau}
&& T_g(\tau)=T_g(\tau_0) \left ( \frac{\tau_0}{\tau} \right )^{r_g-1} \,, \
T_\pi(\tau)=T_\pi(\tau_e) \left ( \frac{\tau_e}{\tau} \right )^{r_\pi-1} \,.
\end{eqnarray}
From these results we obtain the time evolution of the chemical potential
from Eq. (\ref{chemicalpotential}). We find that before the phase transition 
\begin{equation}
\label{cp}
e^{\frac{\mu_g}{T_g}} = e^{\frac{\mu_g(\tau_0)}{T_g(\tau_0)}}
\left [ \frac{T_g}{T_g(\tau_0)} \right ]^{\frac{1}{r_g-1}-3} \,,
\end{equation}
which indicates that for nonzero shear viscosity, $\mu_g$ will decrease to 
be negative during expansion, even if the initial state is in thermal equilibrium
with $\mu_g(\tau_0)=0$. Putting the above relation (\ref{cp}) into the
Gibbs condition (\ref{gibbs}) when the phase transition occurs
\begin{eqnarray}
&&P_g=n_g^c T_c-B=e^{\frac{\mu_c}{T_c}} d_g \frac{T_c^4}{\pi^2} -B
\nonumber \\
&=& P_\pi =n_\pi^c T_c = e^{\frac{\mu_c}{T_c}} d_\pi \frac{T_c^4}{\pi^2} \,,
\end{eqnarray}
we get the transition temperature
\begin{equation}
\label{tc}
T_c=\left \{ e^{-\frac{\mu_g(\tau_0)}{T_g(\tau_0)}} \left [ T_g(\tau_0) \right ]^{\frac{1}{r_g-1}-3}
\frac{\pi^2 B}{d_g-d_\pi} \right \}^{1-\frac{1}{r_g}} \,.
\end{equation}
The chemical potential at the transition temperature, $\mu_c$, can be obtained
from Eq. (\ref{cp}). The dependence of $T_c$ on the initial state is due to
the assumption of the gluon number conservation, which is only valid if the
elastic scatterings are dominant processes. On the other hand,  if inelastic
interactions such like $g+g\leftrightarrow g+g+g$ are as important as the elastic
scatterings, the system will go towards chemical equilibrium, i.e., $\mu_g \to 0$.
The dependence of $T_c$ on the initial state will be almost washed out.
Since it is easier to obtain analytical solutions when considering elastic collisions
only, we do not include inelastic scatterings in the gluonic (and pionic) phase. 

Using $T_g(\tau)$ from Eq. (\ref{temp}) and the value of $T_c$, we obtain
the time $\tau_c$, when the phase transition begins
\begin{equation}
\label{tauc}
\tau_c=\tau_0 \left [ \frac{T_g(\tau_0)}{T_c} \right ]^{1/(r_g-1)} \,.
\end{equation}
Since the time evolutions of the shear viscosity and the entropy density
[see Eq. (\ref{entropydensity})] are known for a chosen constant cross section
$\sigma_g$, the shear viscosity to the entropy density ratio at $\tau_c$ relates
to $\sigma_g$ as
\begin{equation}
\label{eostauc}
\frac{\eta_g(\tau_c)}{s_g(\tau_c)} =\frac{6(d_g-d_\pi)}{5d_g} \frac{T_c^2}{\sigma_g B}
 \frac{1}{ 4-\frac{\mu_g(\tau_0)}{T_g(\tau_0)}-\frac{4-3r_g}{r_g-1} \ln \frac{T_c}{T_g(\tau_0)}}\,.
\end{equation}
We have assumed that $\eta_m/s_m$ is constant during the phase transition.
Therefore, $\eta_m/s_m=\mbox{const.}=\eta_g(\tau_c)/s_g(\tau_c)$. We assume further
that the shear viscosity to the entropy density ratio of the gluonic phase
is same as that of the pionic phase in the mixture. We have then
$\eta_g^c/s_g^c=\eta_\pi^c/s_\pi^c=\mbox{const.}=\eta_m/s_m$.
From Eq. (\ref{eta}) the cross section of pionic scatterings relates to the
cross section of gluonic scatterings as
\begin{equation}
\sigma_\pi= \frac{s_g^c}{s_\pi^c} \sigma_g= \frac{d_g}{d_\pi} \sigma_g \,.
\end{equation} 

We note that the present transport implementation of hadronization with constant
$\eta_m/s_m$ would be equivalent to a hydrodynamic description. The distribution
of hadrons after the dynamic hadronization would be almost the same as that obtained
by using the Cooper-Frye prescription \cite{Cooper:1974mv}, which switches from
viscous hydrodynamic models to hadron transport models
\cite{Luzum:2008cw,Dusling:2009df,Song:2010aq,Schenke:2010rr,Niemi:2011ix}.
However, the assumption of constant $\eta_m/s_m$ made in this article is only for
the comparisons with analytical solutions. In reality the hadronic shear viscosity
to the entropy density ratio may be different from the partonic one, which leads to
a time-dependent $\eta_m/s_m$. The present dynamical scheme of hadronization
provides a possibility to examine the applicability of the Cooper-Frye prescription.

In simulations the initial distribution of gluons at $\tau_0$ is assumed to be thermal
and boost invariant,
\begin{equation}
f(x,p)=\left. e^{-\frac{p^\mu U_\mu}{T_g}} \right |_{\tau_0}=
e^{-\frac{p_\perp \cosh(\bar{\eta}-y)}{T_g(\tau_0)}} \,,
\end{equation}
where $p_\perp$ is the transverse momentum and $\bar{\eta}$ and $y$ are space-time
and momentum rapidity, respectively,
\begin{eqnarray}
\bar{\eta} &=& \frac{1}{2} \ln \frac{t+z}{t-z} \,,\\
y &=& \frac{1}{2} \ln \frac{E+p_z}{E-p_z} \,.
\end{eqnarray}
We consider gluons between a space-time rapidity window $[-\bar{\eta}_M, \bar{\eta}_M]$
with $\bar{\eta}_M=3$. Particles are embedded in a three-dimensional box. The transverse
plane is a $3 \times  3\mbox{ fm}$ square. We use periodical boundary condition
to cancel the transverse expansion. The longitudinal length of the box is set to be long
enough that no particles can exceed the longitudinal bounders at the final time of
observation. The box is equidistantly divided into cells with the same transverse length
$\Delta x=\Delta y$ and the same distance in the space-time rapidity $\Delta \bar{\eta}$.
In simulations we set $\Delta x=\Delta y=0.25 \mbox{ fm}$ and $\Delta \bar{\eta}=0.025$.
To avoid numerical artifacts we use a large value of test particle number, $N_{test}=14000$.

In the following we show how to extract the volume fraction, particle number and
energy density, temperature, and chemical potential of gluons and pions from
the numerical simulation. The particle four-flow and the momentum-energy tensor in
a transverse slice within $\Delta \bar{\eta}$ are calculated by
\begin{eqnarray}
&& N^\mu=\int \frac{d^3p}{(2\pi)^3}\frac{p^\mu}{p^0} f 
=\frac{1}{V_{slice}} \frac{1}{N_{test}} \sum_i \frac{p^\mu_i}{p^0_i} \,, \\
&& T^{\mu\nu}=\int \frac{d^3p}{(2\pi)^3}\frac{p^\mu p^\nu}{p^0} f
=\frac{1}{V_{slice}} \frac{1}{N_{test}} \sum_i 
\frac{p^\mu_i p^\nu_i}{p^0_i} \,,
\end{eqnarray}
where the sums are either over gluons or over pions. From $N_g^\mu$ for gluons
and $N_\pi^\mu$ for pions we calculate the flow velocity $U^\mu$ by using the Eckard's
definition,
\begin{equation}
U^\mu= \frac{N_g^\mu+N_\pi^\mu}{\sqrt{(N_g^\nu+N_\pi^\nu)(N_{g\nu}+N_{\pi\nu})}} \,.
\end{equation}
Then we obtain the particle number and kinetic energy densities in the volume $V_{slice}$
\begin{eqnarray}
\label{density_n}
n'_i&=&N^\mu_i U_\mu \,, \\
\label{density_e}
e'_i&=&U_\mu T^{\mu\nu}_i U_\nu \,,
\end{eqnarray}
where $i$ stands for gluons or pions. The actual densities of gluons and pions are
\begin{eqnarray}
\label{neg}
&&n_g=n'_g/f_g \,, \ \ e_g=e'_g/f_g +B \,,\\
\label{nepi}
&&n_\pi=n'_\pi/(1-f_g) \,, \ \ e_\pi=e'_\pi/(1-f_g) \,.
\end{eqnarray}
We get the temperature of each phases by
\begin{equation}
\label{temp1}
T_g=\frac{e_g-B}{3n_g}=\frac{e'_g}{3n'_g} \,, \ \
T_\pi=\frac{e_\pi}{3n_\pi}=\frac{e'_\pi}{3n'_\pi} \,.
\end{equation} 
Since from $n_i'$ and $e_i'$ we cannot uniquely determine $\mu_g$, $\mu_\pi$, and
$f_g$ in the mixture, we assume that $\mu_g/T_g=\mu_\pi/T_\pi$. Thus, 
\begin{equation}
\frac{n'_g}{n'_\pi} = \frac{f_g n_g}{(1-f_g)n_\pi}=
\frac{f_g}{1-f_g} \frac{n_g^{eq}}{n_\pi^{eq}}
= \frac{f_g}{1-f_g} \frac{d_g T_g^3}{d_\pi T_\pi^3}
\end{equation}
and 
\begin{equation}
\label{fgtau_act}
f_g(\tau) = \left ( 1 + \frac{d_g T_g^3}{d_\pi T_\pi^3} \frac{n'_\pi}{n'_g} \right )^{-1} \,.
\end{equation}
We then have $n_g$, $e_g$, $n_\pi$, and $e_\pi$; see Eqs. (\ref{neg}) and (\ref{nepi}).
From these densities we obtain $\mu_g$ and $\mu_\pi$ according to their definitions
(\ref{chemicalpotential}). 

\section{Numerical results}
\label{sec6:results}
In this section we simulate the phase transition from gluons to pions in a 
one-dimensional expansion with Bjorken boost invariance by implementing the microscopic
processes into the parton cascade model BAMPS.
We will show the numerical results and compare them with the analytical solutions
derived in Secs. \ref{sec4:bjorken} and \ref{sec5:setup}.

As an example, we set the temperature of gluons to be $T_g=0.3 \mbox{ GeV}$ 
at the initial time $\tau_0=0.5 \mbox{ fm/c}$. Since we will compare the numerical
results with the solutions from first-order viscous hydrodynamics, the total cross
section of gluon elastic scatterings is set to be a large value of
$\sigma_g=16.5 \mbox{ mb}$, which leads to a small shear viscosity to the entropy
ratio at the phase transition. With these setups we obtain $T_c=0.2357 \mbox{ GeV}$
and $\tau_c=1.4979 \mbox{ fm/c}$ from Eqs. (\ref{tc}) and (\ref{tauc}). Further we
get $\mu_c/T_c=-0.3735$ and $\eta_g/s_g=0.1045$ at $\tau_c$ from Eqs. (\ref{cp})
and (\ref{eostauc}). In the following we concentrate on the local region at zero
space-time rapidity with a small interval of $0.025$, and calculate densities in
this region.

In the numerical calculation we determine $\tau_c$ as follows. According to
Eqs. (\ref{pg}), (\ref{ppi}), and (\ref{gibbs}) we have $B=(n_g^c-n_\pi^c)T_c$
at the phase transition. With $n_\pi= n_g d_\pi/d_g$, which is only true during
the phase transition, we see that before the phase transition
$(1-d_\pi/d_g) n_g T_g$ is always larger than $B$. Therefore, we get $\tau_c$, once
\begin{equation}
\left ( 1-\frac{d_\pi}{d_g} \right ) n_g(\tau_c) T_g(\tau_c)  < B 
\end{equation}
due to numerical fluctuations at the phase transition. From the simulation we
extract $\tau_c=1.4386 \mbox{ fm/c}$ and accordingly $T_c=0.2269 \mbox{ GeV}$,
which slightly differ from the values expected. Although fluctuations exist in
numerical extractions of $\tau_c$ and $T_c$, the differences from the expected
values have an additional origin. 
\begin{figure}[b]
 \centering
 \includegraphics[width=0.45\textwidth]{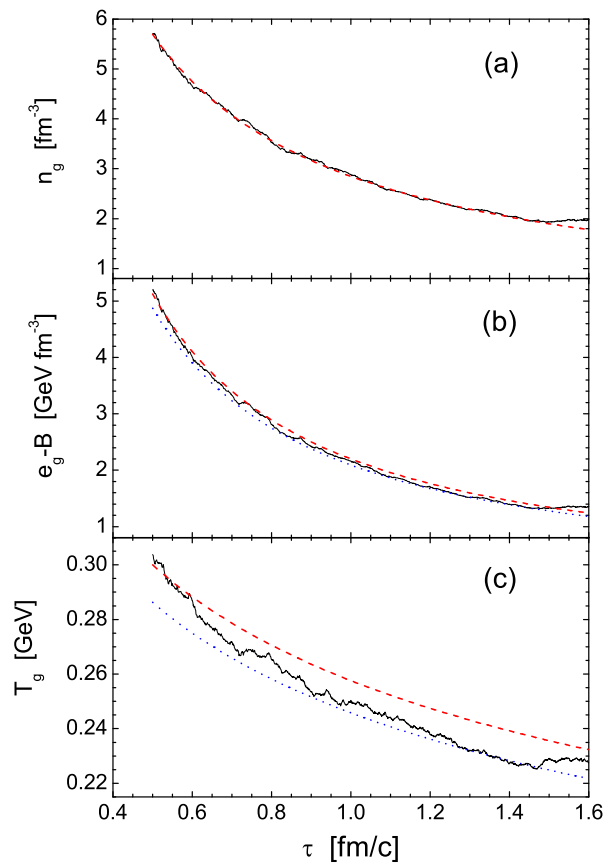}
 \caption{(Color online) The time evolution of the number and kinetic energy
density and the temperature of gluons from $\tau_0=0.5 \mbox{ fm/c}$
to the time shortly after $\tau_c=1.4386 \mbox{ fm/c}$. The numerical results
are depicted by solid curves (in black), while the analytical solutions are shown
by the dashed curves (in red). The dotted curves (in blue) correspond to the shift
of the analytical curves down to meet the values of $e_g$ and $T_g$ at $\tau_c$.}
\label{fig1-net}
\end{figure}
Figure \ref{fig1-net} shows the time evolution
of the number and kinetic energy density and the temperature of gluons
from the initial time $\tau_0$ to the time shortly after $\tau_c$.
We have also depicted the analytical solutions from Eqs. (\ref{ntau}), (\ref{egtau}),
and (\ref{ttau}) by dashed cures. We see a perfect agreement in the number density
$n_g$, as it should be, since we considered only elastic scatterings of gluons
and the time evolution of $n_g$ does not depend on the value of the total cross
section. On the contrary, deviations are visible in the kinetic energy density $e_g-B$
and in the temperature. We shift the analytical curves down to meet the numerical
values of $e_g$ and $T_g$ at $\tau_c$, which correspond to replacing $\tau_0$ by
$\tau_c$ in Eqs. (\ref{egtau}) and (\ref{ttau}). The shifted curves are depicted
by the dotted curves in Fig. \ref{fig1-net}. We see agreements between the shifted
curves and the numerical results from about $1.2 \mbox{ fm/c}$ to $\tau_c$. 
Between $\tau_0$ and $1.2 \mbox{ fm/c}$ we see a relaxation from the thermal initial
condition to the Navier-Stokes state, which has to be described by second-order or
higher order viscous hydrodynamics \cite{Muronga:2003ta,El:2009vj}.

From the simulation we get $\eta_g/s_g=0.1004$ at $\tau_c$, which is slightly
different from the expected value, but agrees with the value, when we use the
shifted curves in Fig. \ref{fig1-net}; i.e., we change $T_g(\tau_0)$ and
$\mu_g(\tau_0)$ in Eq. (\ref{eostauc}) accordingly.

We set $n_g^c=n_g(\tau_c)$, $e_g^c=e_g(\tau_c)$, $s_g^c=s_g(\tau_c)$,
$\mu_c=\mu_g(\tau_c)$, $\eta_m/s_m=\eta_g(\tau_c)/s_g(\tau_c)$,
$n_\pi^c=n_g^c d_\pi/d_g$, $e_\pi^c=(e_g^c-B)d_\pi/d_g$, and
$s_\pi^c= s_g^c d_\pi/d_g$. With these densities extracted at $\tau_c$,
the gluon fraction $f_g$ extracted at $\tau$, and the gluon number $N_g$
extracted at $\tau$ in each cell we compute all the transition
probabilities and the factor $x$ at $\tau$ according to Eqs. (\ref{p22}),
(\ref{p23b}), (\ref{p22b}), and (\ref{px}). Here we employ Eqs. (\ref{hydroexp2})
and (\ref{shearpressure2}) to calculate $\nabla_\mu U^\mu$, $\tilde \pi_g$, and
$\tilde \pi_\pi$ instead of direct extractions from the particle distributions,
in order to avoid numerical uncertainties, which could be reduced by using
larger $N_{test}$. In addition, since $N_g$ is proportional to $N_{test}$ and
the transition probability should be inversely proportional to $N_{test}$
similar to the collision probability in Eq. (\ref{probab}), we multiply all
the transition probabilities by $N_{test}$.

\begin{figure}[b]
 \centering
 \includegraphics[width=0.45\textwidth]{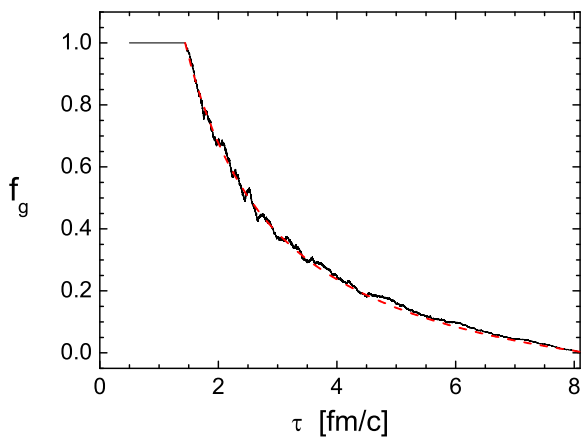}
 \caption{(Color online) The time evolution of the gluon fraction. 
 The solid curve (in black) depicts the numerical result, while the dashed curve
(in red) depicts the expected function.}
 \label{fig2-fg}
\end{figure}

The actual values of $n_g$, $e_g$, $n_\pi$, and $e_\pi$ at $\tau$, calculated
by using Eqs. (\ref{fgtau_act}), (\ref{neg}), and (\ref{nepi}), possess numerical
fluctuations, which induce fluctuations in $T_g$, $T_\pi$, $\mu_g$, $\mu_\pi$,
$s_g$, and $s_\pi$, as seen later in Fig. \ref{fig5-twophase2}. To ensure
$\eta_g/s_g$ and $\eta_\pi/s_\pi$ in the mixture to be equal to $\eta_g/s_g$ at
$\tau_c$, we determine the gluonic (pionic) elastic cross section by
\begin{equation}
\sigma_i(\tau)=\frac{6T_i(\tau)}{5s_i(\tau)} 
\left [ \frac{\eta_g(\tau_c)}{s_g(\tau_c)} \right ]^{-1}
\end{equation}
with $i=g, \pi$ according to Eq. (\ref{eta}). We find (not shown) that the cross
sections during the phase transition fluctuate around the given constant values
in the pure gluonic or pionic phase.

In Fig. \ref{fig2-fg} we compare the numerical extracted gluon fraction $f_g$
[according to Eq. (\ref{fgtau_act})] with the expected function 
[according to Eqs. (\ref{smtau}) and (\ref{fgtau_exp})] and see a perfect agreement.
With the expected function $f_g$ we find the time $\tau_e=8.233 \mbox{ fm/c}$
when the hadronization finishes in the considered volume element. 
Numerically we define $\tau_e$, when on average, the gluon number in a cell
is less than two. We find $\tau_e=8.055 \mbox{ fm/c}$, which is slightly 
earlier than expected. At $\tau_e$ there are still few gluons left (about
1\% of initial gluons), because one gluon in a cell cannot find another gluon
to hadronize. Our numerical handling is as follows: At $\tau_e$ we just rename
the left gluons to pions without any other changes.

\begin{figure}[b]
 \centering
 \includegraphics[width=0.45\textwidth]{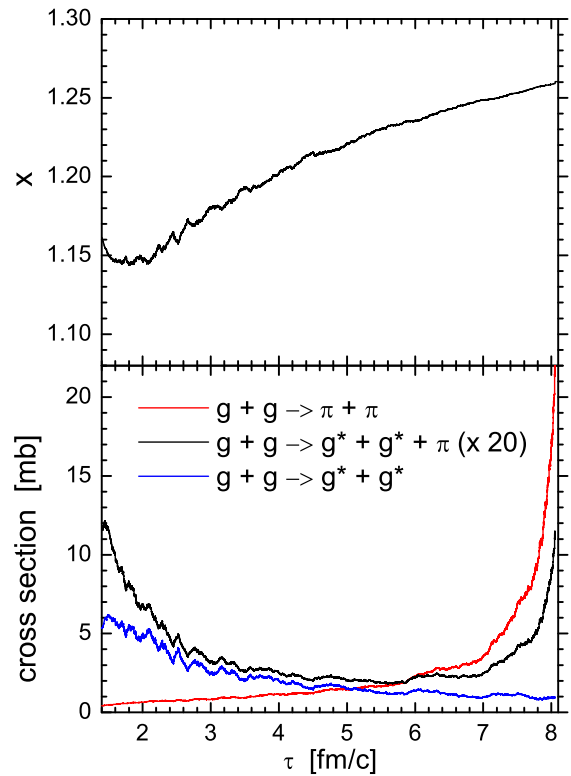}
 \caption{(Color online) The time evolution of the mean transition cross sections
and the factor $x$. }
 \label{fig3-crossection}
\end{figure}   
Analogously to the relation between the collision probability and the cross section in
Eq. (\ref{probab}), we define the transition cross sections of the processes
$g+g\to \pi+\pi$, $g+g\to g^*+g^*+\pi$, and $g+g\to g^*+g^*$ from their transition
probabilities. Figure \ref{fig3-crossection} shows the time evolution of the mean
transition cross sections and the factor $x$ during the phase transition.
The cross section of $g+g\to g^*+g^*+\pi$ is multiplied by $20$ and is negligible
small due to the small value of $\eta_m/s_m$. During the phase transition all
cross sections are below $6 \mbox{ mb}$ except for the cross section of
$g+g\to \pi+\pi$ within $0.5 \mbox{ fm/c}$ before the end of the phase transition,
which increases into infinity. The divergence happens, because shortly
before the complete hadronization the number of gluons is approaching to zero and
on the other hand, the hadronization has an approximately constant rate, i.e.,
$-dV_g \sim A d\tau$, where $A$ is the transverse area. 

\begin{figure}[b]
 \centering
 \includegraphics[width=0.44\textwidth]{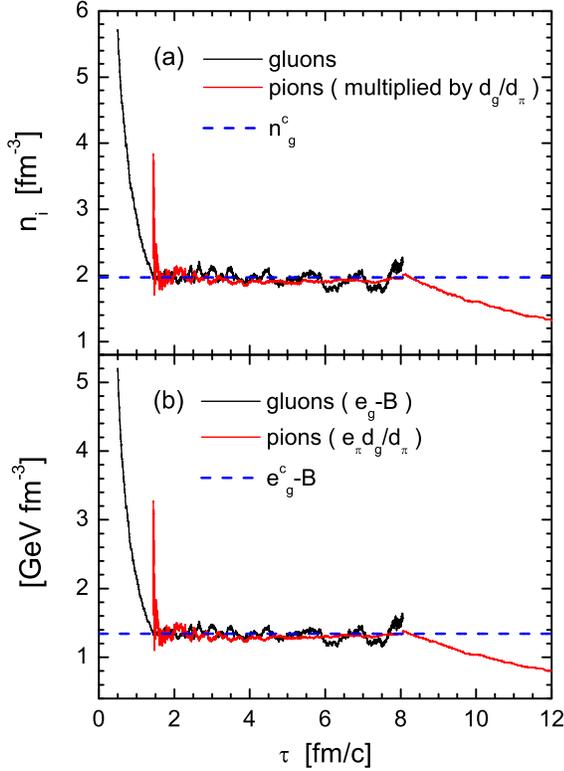}
 \caption{(Color online) The time evolution of the number
and the kinetic energy density. The black (red) curves are for
gluons (pions). The dashed lines depict the values at $\tau_c$.
From $\tau_e=8.055 \mbox{ fm/c}$, the densities of gluons are zero (not plotted).}
 \label{fig4-twophase1}
\end{figure}
Figures \ref{fig4-twophase1}(a) and \ref{fig4-twophase1}(b) show the time evolution
of the number and the kinetic energy density of gluons (black curves) and pions (red curves),
respectively, evaluated according to Eqs. (\ref{neg}), (\ref{nepi}), and (\ref{fgtau_act}).
For comparisons, the densities of pions are multiplied by the ratio of the
degeneracy factors $d_g/d_\pi$. We see good agreements between the gluonic
densities and the amplified pionic densities.
We also see that the densities maintain almost constant during the phase transition,
expect for larger statistical uncertainties of pionic densities after $\tau_c$
and those of gluonic densities before $\tau_e$ due to the small amount of particles.
Both the average values of the gluon number density and the kinetic energy density
agree well with $n_g^c=n_g(\tau_c)=1.9724 \mbox{ fm}^{-3}$ and 
$e_g^c-B=e_g(\tau_c)-B=1.3427 \mbox{ GeV fm}^{-3}$, which are denoted by the
dashed lines.

\begin{figure}[t]
 \centering
 \includegraphics[width=0.45\textwidth]{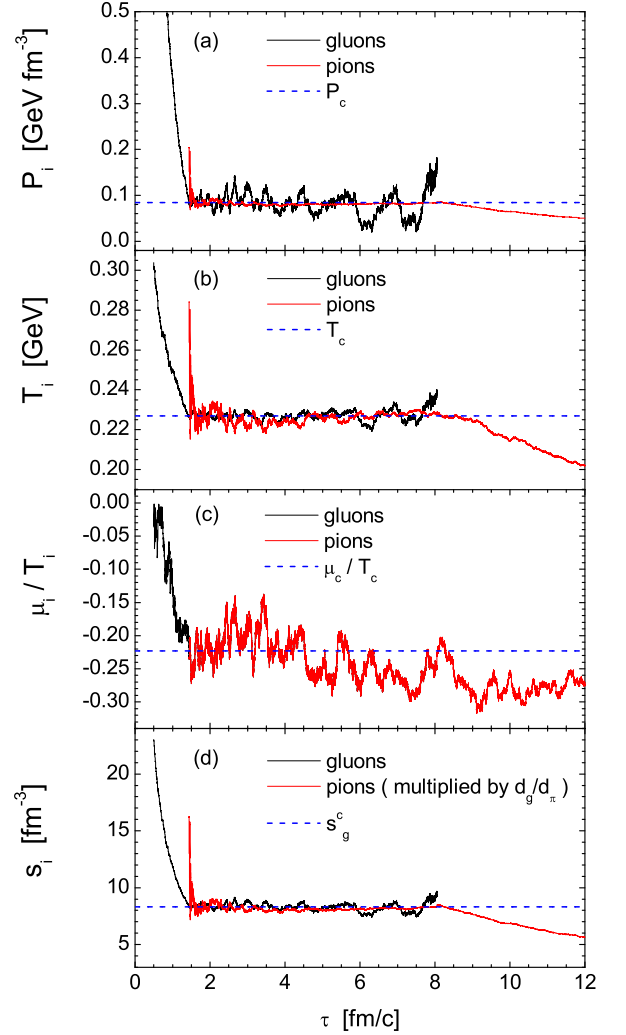}
 \caption{(Color online) Same as Fig. \ref{fig4-twophase1}. From top to bottom:
The time evolution of the pressure, the temperature, the chemical potential to the
temperature ratio, and the entropy density.}
 \label{fig5-twophase2}
\end{figure}
In Fig. \ref{fig5-twophase2} we present the time evolution of the
pressure, temperature, chemical potential, and entropy density, which are obtained
from the number and energy densities shown in Fig. \ref{fig4-twophase1}.
Figure \ref{fig5-twophase2}(a) pictures the pressure of gluons and pions, which are
obtained according to the equations of state Eqs. (\ref{pg}) and (\ref{ppi}).
The temperatures are calculated from Eq. (\ref{temp1}) and shown in Fig. \ref{fig5-twophase2}(b).
We see that the pressures and temperatures are almost constant during the phase transition.
The average values also agree well with $P_c=P_g(\tau_c)=0.0834 \mbox{ GeV fm}^{-3}$
and $T_c=T_g(\tau_c)=0.2269 \mbox{ GeV}$, which are denoted by the dashed lines.
From the number and the kinetic energy density we also obtain the chemical
potential of gluons and pions according to the definition (\ref{chemicalpotential}).
In Fig. \ref{fig5-twophase2}(c) we plot the time evolution of the ratio
of the chemical potential to the temperature. $\mu_g/T_g$ is exactly the same
as $\mu_\pi/T_\pi$ during the phase transition, because this is the assumption
for extracting $f_g$ [see Eq. (\ref{fgtau_act})]. We see that $\mu_g/T_g$
(also $\mu_\pi/T_\pi$) is almost constant around 
$\mu_c/T_c=\mu_g(\tau_c)/T_g(\tau_c)=-0.223$. We have demonstrated that the Gibbs
condition (\ref{gibbs}) is realized in our numerical implementations.
Finally we show in Fig. \ref{fig5-twophase2}(d) the time evolution of the entropy density
of gluons and pions obtained according to Eq. (\ref{entropydensity}).
Same as the number and the kinetic energy density, the entropy density of pions
is multiplied by $d_g/d_\pi$ for comparison. We see that $s_g$ and $s_\pi d_g/d_\pi$
have almost the same constant value during the phase transition. The average value
of $s_g$ agrees well with $s_g^c=s_g(\tau_c)=8.3297 \mbox{ fm}^{-3}$, denoted by
the dashed line.

\begin{figure}[b]
 \centering
 \includegraphics[width=0.44\textwidth]{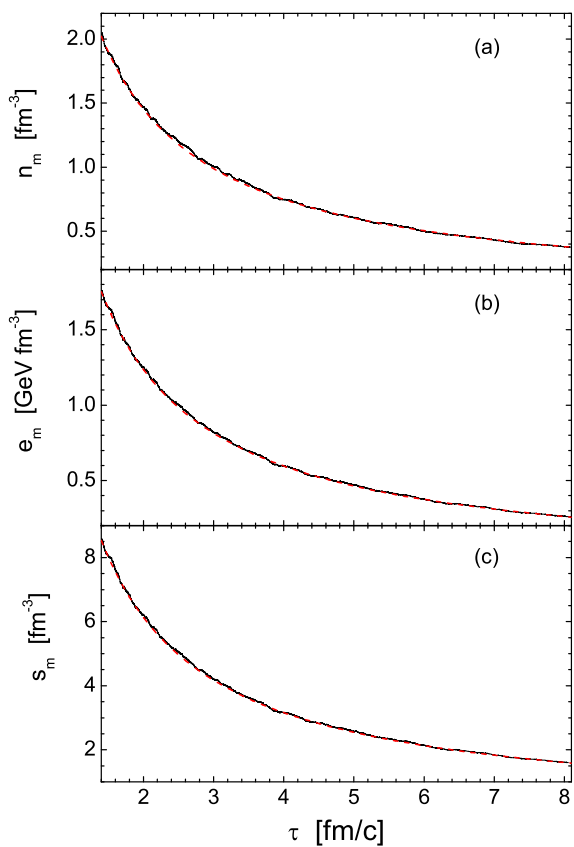}
 \caption{(Color online) The time evolution of the total number, energy, and entropy
density. The solid curves (in black) depict the numerical densities, while the dashed
curves (in red) depict the analytical solutions.}
 \label{fig6-mix}
\end{figure}

After the phase transition is complete in the considered volume element,
the number, energy, and entropy density, and the temperature of pions in
that volume element decrease in time. The numerical results agree well with the
analytical solutions (not shown).  

\begin{figure}[t]
 \centering
 \includegraphics[width=0.45\textwidth]{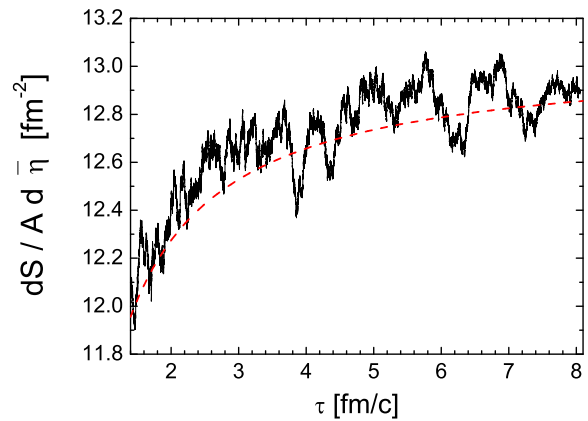}
 \caption{(Color online) Same as Fig. \ref{fig6-mix}, but for the total entropy
 per space-time rapidity per transverse area.}
 \label{fig7-entropy}
\end{figure}

In Fig. \ref{fig6-mix} we present the time evolution of the total number, energy,
and entropy density in the mixed phase according to Eqs. (\ref{nm}), (\ref{em}),
and (\ref{sm}), but replacing $n_g^c$, $n_\pi^c$, $e_g^c$, $e_\pi^c$, $s_g^c$,
and $s_\pi^c$ by the numerical values given in Figs. \ref{fig4-twophase1} and
\ref{fig5-twophase2}.
Comparisons with the analytical solutions given in Eqs. (\ref{nmtau}),
(\ref{emtau}), and (\ref{smtau}) show perfect agreements.
The total entropy per space-time rapidity and per transverse area is obtained
by multiplying the total entropy density by the time $\tau$ and is depicted in
Fig. \ref{fig7-entropy}.
We see that the increase of the total entropy during the hadronization
is realized in our dynamical hadronization scheme and agrees well with the
analytical solution.

\section{Summary and outlook}
\label{sec7:summary}
In this article we have implemented a dynamical hadronization scheme describing 
the first-order confinement and deconfinement phase transition between gluons and pions.
The continuous change of the gluon volume and the pion volume are derived
theoretically by the energy balance according to the condition of the phase equilibrium.
Based on the derived volume changes, the transition probabilities of the considered
microscopic processes $g+g\to \pi+\pi$, $g+g \to \pi+\pi+\pi$, and their back reactions
are determined to mimic the phase transition within a kinetic transport approach.
We have carried out a simulation of the phase transition in a one-dimensional
expansion with Bjorken boost invariance and compared the numerical results
with the analytical solutions. We have seen almost perfect agreements.
This demonstrates the applicability of our dynamical scheme in describing
the first-order confinement and deconfinement phase transition in a more realistic
expansion of the QCD matter produced in relativistic heavy-ion collisions.

In future works we will first improve the present hadronization scheme by adding
quarks and more hadron species and apply it to study the relation between the
collective flow of hadrons and that of quarks and gluons. In particular, we would
like to address the contribution of gluons to the collective flow of hadrons and
to examine whether there is a real quark number scaling. Second, we will
investigate the dissipative effect in the distribution function of hadrons during
the phase transition and quantify the difference from that obtained by using
the Cooper-Frye prescription after viscous hydrodynamic calculations.
Third, we will implement hadronic transport processes and establish a multiphase
transport model, which is able to describe all stages of heavy-ion collisions.
In addition, referring to the dynamics within the chiral $\sigma$ model
\cite{Stephanov:2009ra,Nahrgang:2011mg}
or the Nambu-Jona-Lasinio model \cite{Plumari:2010ah,Marty:2012vs},
we want to include both the confinement and chiral phase transition in one transport
approach, where interactions between particles and fields \cite{Wesp:2014xpa} will
be implemented explicitly.

\section*{Acknowledgement}
ZX thanks P. Huovinen and C. M. Ko for helpful discussions.
This work was financially supported by the NSFC and the MOST under Grants
No. 11275103, No. 11335005, No. 11575092, and No. 2015CB856903. The BAMPS simulations
were performed at Tsinghua National Laboratory for Information Science and Technology.


\begin{thebibliography}{88}
\bibitem{Arsene:2004fa} 
  BRAHMS Collaboration, I.~Arsene {\it et al.},
  Nucl.\ Phys.\ A {\bf 757}, 1 (2005).

\bibitem{Back:2004je} 
  B.~B.~Back {\it et al.},
  Nucl.\ Phys.\ A {\bf 757}, 28 (2005).

\bibitem{Adams:2005dq} 
  STAR Collaboration, J.~Adams {\it et al.},
  Nucl.\ Phys.\ A {\bf 757}, 102 (2005).

\bibitem{Adcox:2004mh} 
  PHENIX Collaboration, K.~Adcox {\it et al.}, 
  Nucl.\ Phys.\ A {\bf 757}, 184 (2005).
  
\bibitem{Aamodt:2010pa} 
  ALICE Collaboration, K.~Aamodt {\it et al.}, 
  Phys.\ Rev.\ Lett.\  {\bf 105}, 252302 (2010).

\bibitem{Aad:2010bu} 
  ATLAS Collaboration, G.~Aad {\it et al.}, 
  Phys.\ Rev.\ Lett.\  {\bf 105}, 252303 (2010).

\bibitem{Chatrchyan:2011sx} 
  CMS Collaboration, S.~Chatrchyan {\it et al.}, 
  Phys.\ Rev.\ C {\bf 84}, 024906 (2011).

\bibitem{Lin:2002rw}
  Z.-W.~Lin and C.~M.~Ko,
  Phys.\ Rev.\ Lett.\  {\bf 89}, 202302 (2002);
  V.~Greco, C.~M.~Ko and P.~Levai,
  {\it ibid.} {\bf 90}, 202302 (2003);
  R.~J.~Fries, B.~M\"uller, C.~Nonaka and S.~A.~Bass,
  {\it ibid.} {\bf 90}, 202303 (2003);
  D.~Molnar and S.~A.~Voloshin,
  {\it ibid.} {\bf 91}, 092301 (2003);
  R.~C.~Hwa and C.~B.~Yang,
  Phys.\ Rev.\  C {\bf 67}, 064902 (2003).

\bibitem{Abelev:2008ae} 
  STAR Collaboration, B.~I.~Abelev {\it et al.}, 
  Phys.\ Rev.\ C {\bf 77}, 054901 (2008).

\bibitem{Luzum:2008cw} 
  M.~Luzum and P.~Romatschke,
  Phys.\ Rev.\ C {\bf 78}, 034915 (2008);
  {\bf 79}, 039903(E) (2009).

\bibitem{Dusling:2009df} 
  K.~Dusling, G.~D.~Moore and D.~Teaney,
  Phys.\ Rev.\ C {\bf 81}, 034907 (2010).

\bibitem{Song:2010aq} 
  H.~Song, S.~A.~Bass and U.~Heinz,
  Phys.\ Rev.\ C {\bf 83}, 024912 (2011).

\bibitem{Schenke:2010rr} 
  B.~Schenke, S.~Jeon and C.~Gale,
  Phys.\ Rev.\ Lett.\  {\bf 106}, 042301 (2011).

\bibitem{Niemi:2011ix} 
  H.~Niemi, G.~S.~Denicol, P.~Huovinen, E.~Molnar and D.~H.~Rischke,
  Phys.\ Rev.\ Lett.\  {\bf 106}, 212302 (2011).

\bibitem{Molnar:2014fva} 
  D.~Molnar and Z.~Wolff,
  arXiv:1404.7850 [nucl-th].

\bibitem{Cooper:1974mv} 
  F.~Cooper and G.~Frye,
  Phys.\ Rev.\ D {\bf 10}, 186 (1974).

\bibitem{Stoecker:2015zea} 
  H.~Stoecker {\it et al.},
  J.\ Phys.\ G {\bf 43}, 015105 (2016).
  
\bibitem{Svetitsky:1982gs} 
  B.~Svetitsky and L.~G.~Yaffe,
  Nucl.\ Phys.\ B {\bf 210}, 423 (1982);
  T.~Celik, J.~Engels and H.~Satz,
  Phys.\ Lett.\ B {\bf 125}, 411 (1983);
  F.~Karsch,
  Nucl.\ Phys.\ A {\bf 698}, 199 (2002);
  S.~Borsanyi, G.~Endrodi, Z.~Fodor, S.~D.~Katz and K.~K.~Szabo,
  J. High Energy Phys. {\bf 1207}, 056 (2012);
  A.~Francis, O.~Kaczmarek, M.~Laine, T.~Neuhaus and H.~Ohno,
  Phys.\ Rev.\ D {\bf 91}, 096002 (2015).

\bibitem{Aoki:2005vt} 
  Y.~Aoki, Z.~Fodor, S.~D.~Katz and K.~K.~Szabo,
  J. High Energy Phys. {\bf 0601}, 089 (2006).

\bibitem{Bazavov:2009zn} 
  A.~Bazavov {\it et al.},
  Phys.\ Rev.\ D {\bf 80}, 014504 (2009).

\bibitem{Xu:2004mz}
  Z.~Xu and C.~Greiner,
  Phys.\ Rev.\  C {\bf 71}, 064901 (2005);
  Z.~Xu and C.~Greiner,
  {\it ibid.} {\bf 76}, 024911 (2007);
  J.~Uphoff, F.~Senzel, O.~Fochler, C.~Wesp, Z.~Xu and C.~Greiner,
  Phys.\ Rev.\ Lett.\  {\bf 114}, 112301 (2015).
  
\bibitem{Lin:2004en} 
  Z.~W.~Lin, C.~M.~Ko, B.~A.~Li, B.~Zhang and S.~Pal,
  Phys.\ Rev.\ C {\bf 72}, 064901 (2005).

\bibitem{Cassing:2008sv} 
  W.~Cassing and E.~L.~Bratkovskaya,
  Phys.\ Rev.\ C {\bf 78}, 034919 (2008).

\bibitem{Chodos:1974je} 
  A.~Chodos, R.~L.~Jaffe, K.~Johnson, C.~B.~Thorn and V.~F.~Weisskopf,
  Phys.\ Rev.\ D {\bf 9}, 3471 (1974).

\bibitem{Xu:2014ega} 
  Z.~Xu, K.~Zhou, P.~Zhuang and C.~Greiner,
  Phys.\ Rev.\ Lett.\  {\bf 114}, 182301 (2015).

\bibitem{Rischke:1995mt} 
  D.~H.~Rischke, Y.~Pursun and J.~A.~Maruhn,
  Nucl.\ Phys.\ A {\bf 595}, 383 (1995);
  {\bf 596}, 717(E) (1996).
  
\bibitem{Sollfrank:1996hd} 
  J.~Sollfrank, P.~Huovinen, M.~Kataja, P.~V.~Ruuskanen, M.~Prakash and R.~Venugopalan,
  Phys.\ Rev.\ C {\bf 55}, 392 (1997).

\bibitem{Kolb:2000sd} 
  P.~F.~Kolb, J.~Sollfrank and U.~W.~Heinz,
  Phys.\ Rev.\ C {\bf 62}, 054909 (2000).

\bibitem{Chomaz:2003dz} 
  P.~Chomaz, M.~Colonna and J.~Randrup,
  Phys.\ Rept.\  {\bf 389}, 263 (2004).

\bibitem{Steinheimer:2012gc} 
  J.~Steinheimer and J.~Randrup,
  Phys.\ Rev.\ Lett.\  {\bf 109}, 212301 (2012).

\bibitem{Li:2016uvu} 
  F.~Li and C.~M.~Ko,
  arXiv:1606.05012 [nucl-th].

\bibitem{Muronga:2003ta} 
  A.~Muronga,
  Phys.\ Rev.\ C {\bf 69}, 034903 (2004).

\bibitem{Bjorken:1982qr} 
  J.~D.~Bjorken,
  Phys.\ Rev.\ D {\bf 27}, 140 (1983).

\bibitem{Huovinen:2008te} 
  P.~Huovinen and D.~Molnar,
  Phys.\ Rev.\ C {\bf 79}, 014906 (2009).
  
\bibitem{Wesp:2011yy} 
  C.~Wesp, A.~El, F.~Reining, Z.~Xu, I.~Bouras and C.~Greiner,
  Phys.\ Rev.\ C {\bf 84}, 054911 (2011).

\bibitem{El:2012cr} 
  A.~El, F.~Lauciello, C.~Wesp, Z.~Xu and C.~Greiner,
  Nucl.\ Phys.\ A {\bf 925}, 150 (2014).

\bibitem{El:2009vj} 
  A.~El, Z.~Xu and C.~Greiner,
  Phys.\ Rev.\ C {\bf 81}, 041901 (2010).
 
\bibitem{Stephanov:2009ra} 
  M.~A.~Stephanov,
  Phys.\ Rev.\ D {\bf 81}, 054012 (2010).

\bibitem{Nahrgang:2011mg} 
  M.~Nahrgang, S.~Leupold, C.~Herold and M.~Bleicher,
  Phys.\ Rev.\ C {\bf 84}, 024912 (2011).
  
\bibitem{Plumari:2010ah} 
  S.~Plumari, V.~Baran, M.~Di Toro, G.~Ferini and V.~Greco,
  Phys.\ Lett.\ B {\bf 689}, 18 (2010).
  
\bibitem{Marty:2012vs} 
  R.~Marty and J.~Aichelin,
  Phys.\ Rev.\ C {\bf 87}, 034912 (2013).

\bibitem{Wesp:2014xpa} 
  C.~Wesp, H.~van Hees, A.~Meistrenko and C.~Greiner,
  Phys.\ Rev.\ E {\bf 91}, 043302 (2015);
  C.~Greiner, C.~Wesp, H.~van Hees and A.~Meistrenko,
  J.\ Phys.\ Conf.\ Ser.\  {\bf 636}, 012007 (2015).

\end{thebibliography}
\end{document}